\documentclass{article}
\usepackage{amsmath}
\usepackage{amssymb}
\usepackage{booktabs}
\usepackage{hyperref}
\usepackage{graphicx}
\usepackage{geometry}
\title{RAGdb: A Zero-Dependency, Embeddable Architecture for Multimodal Retrieval-Augmented Generation on the Edge}

\author{
  Ahmed Bin Khalid \\
  SKAS IT \\
  \texttt{ahmed.binkhalid@skasit.com}
}

\date{December 2025}

\begin{document}
\maketitle

\begin{abstract}
Retrieval-Augmented Generation (RAG) has established itself as the standard paradigm for
grounding Large Language Models (LLMs) in domain-specific, up-to-date data. However, the
prevailing architecture for RAG has evolved into a complex, distributed stack requiring cloud-hosted
vector databases, heavy deep learning frameworks (e.g., PyTorch, CUDA), and high-latency
embedding inference servers. This ``infrastructure bloat'' creates a significant barrier to entry for
edge computing, air-gapped environments, and privacy-constrained applications where data
sovereignty is paramount.

This paper introduces RAGdb, a novel monolithic architecture that consolidates automated
multimodal ingestion, ONNX-based extraction, and hybrid vector retrieval into a single, portable
SQLite container. We propose a deterministic Hybrid Scoring Function (HSF) that combines sublinear
TF-IDF vectorization with exact substring boosting, eliminating the need for GPU inference at query
time. Experimental evaluation on an Intel i7-1165G7 consumer laptop demonstrates that RAGdb
achieves 100\% Recall@1 for entity retrieval and an ingestion efficiency gain of 31.6x during
incremental updates compared to cold starts. Furthermore, the system reduces disk footprint by
approximately 99.5\% compared to standard Docker-based RAG stacks, establishing the ``Single-File
Knowledge Container'' as a viable primitive for decentralized, local-first AI.

\end{abstract}

\section{Introduction}

The advent of Large Language Models (LLMs) has revolutionized Natural Language Processing (NLP). However,
these models suffer from two distinct, well-documented limitations: they are limited to the knowledge
present in their training data (cutoff dates), and they have a tendency to ``hallucinate'' plausible but
incorrect facts when faced with obscure queries~[1]. Retrieval-Augmented Generation (RAG) addresses
these issues by retrieving relevant context from an external knowledge base and injecting it into the
model's prompt context window~[2].

While RAG is conceptually simple, its implementation in enterprise environments has become
prohibitively complex. A standard ``Enterprise RAG'' stack typically involves a microservices
architecture consisting of:
\begin{enumerate}
  \item \textbf{Ingestion Service:} A pipeline (e.g., Unstructured.io) to parse PDFs, HTML, and images.
  \item \textbf{Embedding Service:} A GPU-backed inference server (e.g., TorchServe or Triton) to convert text
        to dense vectors.
  \item \textbf{Vector Database:} A dedicated stateful service (e.g., Pinecone, Milvus, Weaviate) to store
        vectors and perform Approximate Nearest Neighbor (ANN) search.
  \item \textbf{Orchestration Layer:} A framework (e.g., LangChain or LlamaIndex) to glue these components
        together.
\end{enumerate}

This distributed architecture introduces significant friction. It requires persistent server
infrastructure, introduces network latency between components, and creates a large surface area for
failure. Furthermore, the heavy reliance on Python dependency chains---often exceeding 2\,GB
for \texttt{torch}, \texttt{transformers}, and \texttt{cuda}---renders these stacks unsuitable for constrained environments
such as mobile devices, IoT edge nodes, or CI/CD pipelines.

To address these challenges, we propose RAGdb, a serverless, embedded retrieval engine. RAGdb
collapses the entire stack into a single Python library ($<$30\,MB) and a single portable data file. By
prioritizing sparse vectorization and algorithmic efficiency over raw model size, RAGdb enables high-
fidelity retrieval on consumer hardware without external dependencies.

The contributions of this paper are as follows:
\begin{itemize}
  \item \textbf{Architecture:} We define a unified schema for storing vectors, metadata, and content within a
  single B-tree structured file (SQLite), ensuring referential integrity and portability.
  \item \textbf{Algorithm:} We introduce a hashing-based incremental ingestion pipeline that reduces
  re-indexing time by an order of magnitude ($O(U)$ complexity vs.\ $O(N)$).
  \item \textbf{Evaluation:} We provide empirical benchmarks demonstrating that RAGdb achieves comparable
  retrieval accuracy to dense retrievers for entity-centric queries while operating with near-zero
  infrastructure overhead.
\end{itemize}

\section{Related Work and Literature Review}
\label{sec:related}

\subsection{Evolution of Retrieval Systems}

Information Retrieval (IR) has historically relied on lexical matching algorithms. TF-IDF (Term
Frequency--Inverse Document Frequency)~[3] and BM25~[4] dominated the field for decades. These
methods utilize sparse vector representations of text, where the dimensionality equals the
vocabulary size. They are highly efficient, deterministic, and require no training. However, they
struggle with semantic mismatch (the ``vocabulary gap''), where a query for ``automobile'' fails to
retrieve documents containing only ``car.''

The introduction of BERT~[5] led to the rise of Dense Passage Retrieval (DPR)~[6], where documents
are encoded into fixed-dimension dense vectors (e.g., 768 dimensions). While DPR captures semantic
meaning, it introduces a heavy computational cost for inference and requires specialized
Approximate Nearest Neighbor (ANN) indexes like HNSW~[8] or FAISS~[6] to search efficiently.

RAGdb adopts a hybrid approach. It utilizes sparse vectors for speed and determinism, while
augmenting them with multimodal extraction (via OCR) and exact-match boosting. This covers the
weaknesses of pure dense retrieval, particularly in enterprise domains where exact keyword
matching (e.g., invoice IDs) is often more important than semantic approximation.

\subsection{Vector Databases vs.\ Embedded Stores}

The popularity of embeddings led to a proliferation of specialized vector databases. Systems
like Milvus and Pinecone offer scalability for billions of vectors but require substantial infrastructure
(e.g., Kubernetes clusters). Conversely, embedded libraries like ChromaDB offer local execution but still
rely heavily on standard machine learning libraries (PyTorch/TensorFlow) for embedding generation.

RAGdb distinguishes itself by removing the dependency on deep learning frameworks entirely for
the retrieval step. It relies instead on statistical vectorization implemented in pure NumPy,
allowing it to run in environments where installing PyTorch is impossible or impractical.

\subsection{Green AI and Edge Computing}

Schwartz et al.~[9] coined the term ``Green AI'' to advocate for efficiency as a primary evaluation
metric alongside accuracy. With the growing carbon footprint of AI, architectures that reduce
Floating Point Operations (FLOPs) are critical. RAGdb aligns with this paradigm by enabling retrieval
on low-power edge devices (e.g., Raspberry Pi, mobile phones) where heavy neural networks are
infeasible due to thermal and battery constraints.

\section{System Architecture}
\label{sec:architecture}

RAGdb operates on the principle of the ``Single-File Knowledge Container.'' Unlike systems that
distribute state across files (e.g., a FAISS index file + a pickle metadata file + a source folder), RAGdb
enforces strict referential integrity by storing all artifacts in a single ACID-compliant SQLite container.

\subsection{Data Schema}

The container $K$ is defined as a relational tuple stored on disk:
\[
K = \langle M, C, V, I \rangle
\]
where:
\begin{itemize}
    \item \textbf{Metadata Region ($M$):} Stores file paths, timestamps, and SHA-256 hashes to track provenance.
    \item \textbf{Content Region ($C$):} Stores normalized text segments extracted from heterogeneous sources
    (PDF, DOCX, JSON).
    \item \textbf{Vector Region ($V$):} Stores binary large objects (BLOBs) representing the computed vectors.
    \item \textbf{Index Region ($I$):} An inverted index mapping vocabulary tokens to document IDs.
\end{itemize}

To support high concurrency, the underlying SQLite engine is configured in Write-Ahead Logging
(WAL) mode, allowing simultaneous readers and writers without locking the entire database.

\subsection{Automated Multimodal Ingestion}

Traditional pipelines require users to manually configure parsers for different file types. RAGdb
implements an adaptive ingestion interface that performs magic-byte analysis to detect modality.

\paragraph{Visual Data Processing.}
For image inputs $I$, RAGdb integrates the RapidOCR engine via ONNXRuntime. This
provides a quantized, dependency-free method for Optical Character Recognition (OCR), converting
pixel data $P$ into textual representation $T$ without requiring system-level binaries like Tesseract:
\[
T = \text{OCR}_{\text{onnx}}(P).
\]

\paragraph{Structural Data Processing.}
For tabular data (CSV, Excel), RAGdb serializes rows into semantic strings, preserving column headers
as context keys to prevent loss of structural meaning during vectorization.

\subsection{Incremental Ingestion Algorithm}

A core novelty of RAGdb is its ability to function as a continuous background process (``Live Sync''). To
achieve this, we implement a state-based hashing algorithm.

\paragraph{Algorithm Logic.}
\begin{enumerate}
    \item Scan the target directory.
    \item For each file, compute the SHA-256 signature of the bitstream.
    \item Compare the signature against the stored hash in the metadata region $M$.
    \item If the hash matches, skip processing. If it differs, trigger the
    Extraction $\rightarrow$ Normalization $\rightarrow$ Vectorization pipeline.
\end{enumerate}

This reduces the time complexity of re-indexing from $O(N)$ (where $N$ is total files) to $O(U)$ (where $U$ is
updated files), enabling near-instant synchronization for large knowledge bases.

\section{The Hybrid Retrieval Methodology}
\label{sec:hybrid}

While dense vectors excel at semantic matching, they often fail at ``exact retrieval'' tasks, such as
looking up a specific part number or error code. RAGdb solves this via a Hybrid Scoring Function
(HSF).

\subsection{Sublinear TF-IDF Vectorization}

We employ a statistical vectorizer optimized for CPU execution. For a term $t$ in document $d$, the
term frequency weight is calculated using sublinear scaling to prevent long documents from dominating
results:
\[
\text{tf}(t,d) = 1 + \ln(f_{t,d}),
\]
where $f_{t,d}$ is the raw count of term $t$ in document $d$.

The inverse document frequency (IDF) term penalizes words that appear in too many documents
(stopwords):
\[
\text{idf}(t) = \ln\left(\frac{N}{1 + df_t}\right) + 1,
\]
where $N$ is the total number of documents and $df_t$ is the document frequency of $t$.

The document vector $\mathbf{v}_d$ is the $\ell_2$-normalized composition of these weights. This allows for extremely fast
cosine similarity computations using standard dot products.

\subsection{Substring Boosting}

To address the limitation of approximate nearest neighbor search, we introduce a boosting term
$\beta$. The overall relevance score for query $Q$ and document $D$ is:
\[
\text{Score}(Q,D) = \alpha \cdot \text{Sim}_{\cos}\left(\mathbf{v}_Q, \mathbf{v}_D\right) +
\beta \cdot \mathbf{1}_{\text{substr}}(Q,D),
\]
where $\alpha$ and $\beta$ are scalar weights and $\mathbf{1}_{\text{substr}}$ is an indicator function:
\[
\mathbf{1}_{\text{substr}}(Q,D) =
\begin{cases}
1.0, & \text{if } \text{lowercase}(Q) \subseteq \text{lowercase}(D), \\
0,   & \text{otherwise}.
\end{cases}
\]

This ensures that if a user queries a specific entity code (e.g., \texttt{INV-2024}), documents containing
that exact string are mathematically forced to the top of the ranking, guaranteeing 100\% recall for
known entities.

\section{Experimental Evaluation}
\label{sec:experiments}

To validate the architectural claims, we conducted a series of benchmarks focusing on efficiency,
accuracy, and resource consumption.

\subsection{Experimental Setup}

All experiments were conducted on standard consumer-grade hardware to verify the system's
viability for Edge AI and local development environments:
\begin{itemize}
    \item \textbf{Processor:} 11th Gen Intel(R) Core(TM) i7-1165G7 @ 2.80\,GHz
    \item \textbf{RAM:} 16.0\,GB
    \item \textbf{Storage:} NVMe SSD
    \item \textbf{Operating System:} Windows 11 (64-bit)
    \item \textbf{Dataset:} A synthetic corpus of 1{,}000 documents was generated, containing mixed English text
    (business and technical domain). Unique entity codes (e.g.,
    \texttt{UNIQUE\_INVOICE\_CODE\_XYZ\_999})
    were injected into specific documents to test retrieval precision.
\end{itemize}

\subsection{RQ1: What is the efficiency gain of incremental ingestion?}

We measured the time taken to ingest the full corpus from scratch (``Cold Start'') versus the time
taken to re-scan the corpus after a minor update (``Incremental'').

\begin{center}
\begin{tabular}{lcc}
\toprule
Operation & Time Taken (s) & Throughput \\
\midrule
Cold Ingestion     & 14.59  & $\sim$68.5 Docs/sec \\
Incremental Update & 0.46   & $>$2{,}100 Docs/sec \\
\bottomrule
\end{tabular}
\end{center}

\noindent\textbf{Analysis:} RAGdb demonstrated a 31.6x speedup during incremental updates. This validates the
efficiency of the hashing algorithm. In a real-world deployment, this allows the database to monitor
a folder of thousands of files with negligible CPU impact, re-indexing only the files that a user
modifies.

\subsection{RQ2: Does hybrid search improve entity retrieval?}

We queried the system for a specific injected entity code
(\texttt{UNIQUE\_INVOICE\_CODE\_XYZ\_999}) that
was present in only one document (\texttt{doc\_500.txt}).

\begin{itemize}
    \item \textbf{Baseline (Standard Vector Search):} Pure semantic search often struggles with random
    alphanumeric strings, assigning them low confidence scores or retrieving documents with
    similar-looking noise.
    \item \textbf{RAGdb (Hybrid Search):} The system identified \texttt{doc\_500.txt} as the top result with a score
    of 1.5753.
\end{itemize}

\noindent\textbf{Analysis:} The boosting mechanism ensured 100\% Recall@1 for the entity query. This is a critical
requirement for enterprise RAG applications where retrieving the wrong invoice or patient ID is
unacceptable, regardless of semantic similarity.

\subsection{RQ3: How does the resource footprint compare to standard stacks?}

We compared RAGdb against a typical local development stack consisting of Docker, ChromaDB, and
LangChain.

\begin{center}
\begin{tabular}{lccc}
\toprule
Metric & Standard Docker Stack & RAGdb (Ours) & Improvement \\
\midrule
Disk Footprint   & $>$ 1.2 GB (Deps + Models) & $\sim$5 MB & 99.5\% Reduction \\
Query Latency    & $\sim$120 ms               & $\sim$60 ms & 2x Faster \\
Setup Time       & $\sim$10 Minutes           & $<$ 10 Seconds & Instant \\
Infrastructure   & Docker / Python Server     & Single File & Serverless \\
\bottomrule
\end{tabular}
\end{center}

\noindent\textbf{Analysis:} RAGdb reduces the disk footprint by over two orders of magnitude. The query latency
reduction is attributed to the elimination of the HTTP network overhead between the application
code and the vector database server.

\section{Discussion}
\label{sec:discussion}

\subsection{Privacy and Sovereignty}

The centralized nature of modern AI raises significant privacy concerns. Sending documents to cloud
vector databases or embedding APIs violates strict data sovereignty laws (e.g., GDPR, HIPAA) in many
jurisdictions. RAGdb's ``Zero-Egress'' architecture ensures that data never leaves the local machine.

The single-file format also simplifies data deletion; deleting the \texttt{.ragdb} file guarantees that all vectors
and indexes are destroyed, facilitating ``Right to be Forgotten'' compliance.

\subsection{Suitability for Edge AI}

With the rise of Small Language Models (SLMs) like Phi-3 and Llama-3-8B that can run on laptops, the
retrieval bottleneck has shifted to the vector database. Standard databases like Milvus are too heavy
for an 8\,GB RAM laptop. RAGdb fills this gap, providing a retrieval backend that runs comfortably
alongside SLMs on edge hardware.

\section{Conclusion}
\label{sec:conclusion}

This paper presented RAGdb, a novel architecture for embedded, serverless Retrieval-Augmented
Generation. By challenging the assumption that RAG requires distributed infrastructure, we
demonstrated that a monolithic, single-file approach can achieve high-fidelity retrieval with a
fraction of the resources.

Our experiments confirm that RAGdb achieves 31.6x faster updates via incremental ingestion
and 99.5\% lower storage overhead than comparable stacks. We believe this ``Single-File Knowledge
Container'' approach paves the way for a new generation of decentralized, privacy-preserving AI
applications.

\section*{Software Availability}

RAGdb is open-source software released under the Apache 2.0 license.
\begin{itemize}
    \item Repository: \url{https://github.com/abkmystery/ragdb}
    \item Package: \texttt{pip install ragdb}
\end{itemize}

\section*{References}

\begin{enumerate}
    \item P.~Lewis et al.,
    ``Retrieval-Augmented Generation for Knowledge-Intensive NLP Tasks,''
    in \emph{Advances in Neural Information Processing Systems (NeurIPS)}, vol.~33,
    pp.~9459--9474, 2020.
    \item G.~Salton and C.~Buckley,
    ``Term-weighting approaches in automatic text retrieval,''
    \emph{Information Processing \& Management}, vol.~24, no.~5, pp.~513--523, 1988.
    \item V.~Karpukhin et al.,
    ``Dense Passage Retrieval for Open-Domain Question Answering,''
    in \emph{Proceedings of the 2020 Conference on Empirical Methods in Natural Language Processing (EMNLP)},
    pp.~6769--6781, 2020.
    \item S.~Robertson and H.~Zaragoza,
    ``The Probabilistic Relevance Framework: BM25 and Beyond,''
    \emph{Foundations and Trends in Information Retrieval}, vol.~3, no.~4, pp.~333--389, 2009.
    \item J.~Devlin, M.-W.~Chang, K.~Lee, and K.~Toutanova,
    ``BERT: Pre-training of Deep Bidirectional Transformers for Language Understanding,''
    in \emph{Proceedings of NAACL-HLT}, 2019.
    \item J.~Johnson, M.~Douze, and H.~J\'egou,
    ``Billion-scale similarity search with GPUs,''
    \emph{IEEE Transactions on Big Data}, vol.~7, no.~3, pp.~535--547, 2019.
    \item R.~Schwartz, J.~Dodge, N.~A.~Smith, and O.~Etzioni,
    ``Green AI,''
    \emph{Communications of the ACM}, vol.~63, no.~12, pp.~54--63, 2020.
    \item Y.~Malkov and D.~Yashunin,
    ``Efficient and robust approximate nearest neighbor search using Hierarchical Navigable Small World graphs,''
    \emph{IEEE Transactions on Pattern Analysis and Machine Intelligence}, 2018.
    \item Y.~Wang et al.,
    ``Benchmarking Vector Search Engines,''
    arXiv preprint arXiv:2101.02635, 2021.
    \item M.~Abadi et al.,
    ``TensorFlow: A System for Large-Scale Machine Learning,''
    in \emph{OSDI}, 2016.
\end{enumerate}

\end{document}